\documentclass{aa}  

\usepackage{natbib}
\usepackage{threeparttable}
\usepackage{diagbox}
\usepackage{slashbox}
\usepackage{multirow}
\usepackage{float}
\usepackage{booktabs}
\usepackage{graphicx,subfigure}
\usepackage{txfonts}
\usepackage[]{hyperref}
\usepackage{appendix}
\newcommand{\urlNewWindow}[1]{\href[pdfnewwindow=true]{#1}{\nolinkurl{#1}}}
\usepackage{color}
\newcommand{\G}{\mathcal{G}}

\usepackage{epstopdf}

\begin{document} 
  \title{Star-planet tidal interaction and the limits of gyrochronology}
  \author{F. Gallet\inst{1} \and P. Delorme\inst{1} 
          }

\offprints{F. Gallet,\\ email: florian.gallet1@univ-grenoble-alpes.fr}

  \institute{$^1$ Univ. Grenoble Alpes, CNRS, IPAG, 38000 Grenoble, France \\
             }

  \date{Received -- ; Accepted --}

  \abstract
{Age estimation techniques such as gyrochronology and magnetochronology cannot be applied to stars that have exchanged angular momentum with their close environments. This is especially true for a massive close-in planetary companion  (with a period  of a few days or less) that could have been strongly impacted by the rotational evolution of the host star, throughout the stellar evolution, through the star-planet tidal interaction.}
{{In this article, we  provide the community with a reliable region in which empirical techniques such as gyrochronology can be used with confidence}.}
{We combined a stellar angular momentum evolution code with a planetary orbital evolution code to study in detail the impact of star-planet  tidal interaction on the evolution of the surface rotation rate of the star.}
{{We show  that the interaction of a close-in massive planet with its host star can strongly modify the surface rotation rate of this latter, in most of the cases associated with a planetary engulfment. A modification of the surface rotation period of more than 90\% can survive a few hundred  Myr after the event and a modification of  10\% can last for a few Gyr. In such cases, a gyrochronology analysis of the star would incorrectly make it appear as rejuvenated, thus preventing us from using this method with confidence. To try overcome this issue, we  proposed the proof of concept of a new age determination technique that we call the tidal-chronology method, which is based on the observed pair $\rm P_{rot,\star}$-$\rm P_{orb}$ of a given star-planet system, where $\rm P_{rot,\star}$ is the stellar surface rotational period and $\rm P_{orb}$ the planetary orbital period.}}
{The gyrochronology technique can only be applied to isolated stars or star-planet systems outside a specific range of $\rm P_{rot,\star}$-$\rm P_{orb}$. This region tends to expand for increasing stellar and planetary mass. In that forbidden region, or if any planetary engulfment is suspected, gyrochronology should be used with extreme caution, while tidal-chronology could be considered. This technique does not provide a precise age for the system yet; however, it {is already an extension of gyrochronology and} could be helpful to determine a more precise range of possible ages for planetary systems composed of a star between 0.3 and 1.2 $M_{\odot}$ and a planet more massive than 1 $\rm M_{jup}$ initially located at a few hundredths of au from the host star.}



\maketitle


\section{Introduction}
Determining the age of a star and in substance that of a planetary system is of prime importance as it provides information about the characteristic timescale of planet formation and migration that can strongly constrain ongoing planetary models \citep{Ida08,Mordasini09,Mordasini12,Alibert13}. {It can also be used to constrain the star-planet interaction efficiency \citep{Lanza11}, {the nature of the star (its internal structure and chemical composition) based on stellar models,} and the stellar phases the planets were subject to during their evolution \citep{Gallet17}.}
\begin{figure*}[!ht]
    \begin{center}
            \includegraphics[width=\linewidth]{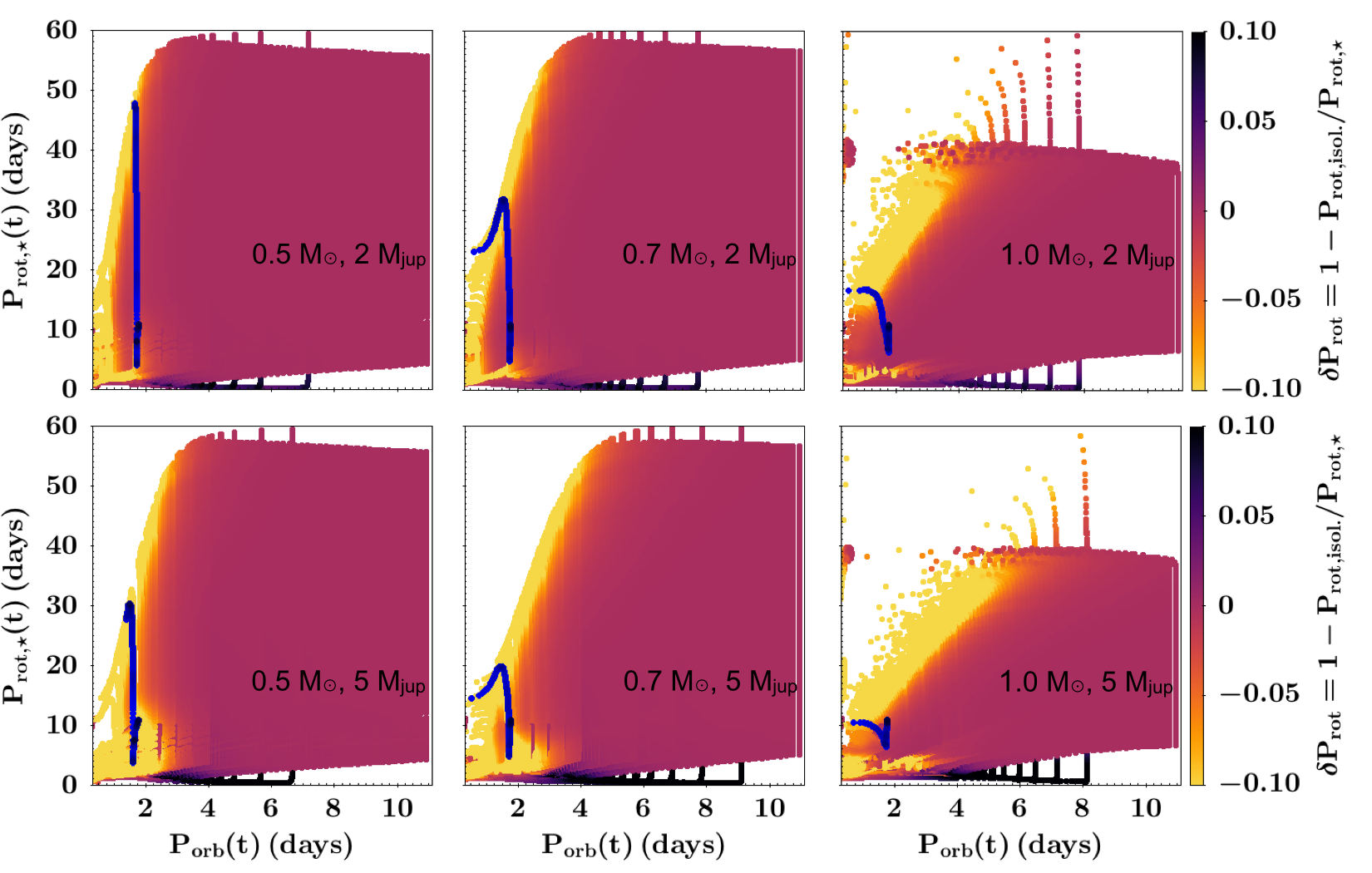}
        \caption{{Evolution of $\rm P_{rot,\star} (t)$ as a function of $\rm P_{orb}(t)$ and stellar mass for a 2 $\rm M_{jup}$ (upper panel) and 5 $\rm M_{jup}$ (lower panel)  planet. The color gradient corresponds to the {relative} rotational departure $\rm \delta P_{rot}=1 - P_{rot,isol.}/P_{rot,\star}$. Only systems where the planet is still present are plotted. {The blue lines (composed of plain circles) depict the evolution of a given star-planet system ($\rm P_{rot,init}=10.8$ days and $\rm SMA_{init}=30\% R_{co}$) in the $\rm P_{rot,\star} (t)-\rm P_{orb} (t)$ plane.}} The red area shows the domain of applicability of gyrochronology analysis.}
        \label{DeltaP3D}%
    \end{center}
\end{figure*}
{There are three categories of age determination technique   (see \citealt{Soderblom10,Soderblom14} and the review of \citealt[][]{Jeffries14} from the 2013 Evry Schatzman school)}: 
\begin{itemize}
        \item Semi-fundamental techniques, which rely on age-dependant phenomena that are physically well understood (e.g., nucleocosmochronometry based on the decay of long lived isotope, as  is used for the Sun);
        \item Model-dependant techniques,  such as isochrone fitting in the Hertzsprung-Russell diagram (HRD), asteroseismology, and white dwarf cooling,  where the physics is mostly understood, but part of it is simplified or described using tuning parameters (e.g., the mixing length theory);
        \item Empirical techniques, where the understanding of the physics is qualitative and described using free parameters that are constrained using observations. They are based on the calibration of age-dependant phenomena using similar observations of stars of known age, which are then used to estimate the ages of other stars \citep[e.g., the gyrochronology technique, see][]{Barnes07}.
\end{itemize}
In  the case of isolated stars (i.e., stars not impacted by interaction with the close environment during {their} evolution), their ages can be determined by putting them in a  color-magnitude diagram (CMD) and fitting the observed sequence with stellar isochrone models. This technique obviously depends upon the stellar model used and can result in discrepancies from one model to another \citep{Lebreton14}. Many systematics still exist, such as the effect of metallicity, the magnetic field, and the stellar model hypothesis used to produce the isochrones employed for the age determination, which prevent us from providing reliable age estimates. Additionally, to apply this technique, precise distance measurements are  required in order to accurately determine the stellar parameters (i.e., luminosity and magnitudes in given bands). The Gaia mission \citep{Gaia,DR2} is revolutionizing this field by providing the community with  parallax measurements of unprecedented high quality. 

Empirical techniques such as gyrochronology \citep{Barnes03} and magnetochronology \citep{Vidotto14} also exist. Both are based on the observation that during the main-sequence  (MS) phase the evolution of the surface rotation and magnetic field of a star, for a given mass, only depend on age. {Provided that the stellar rotation period (respectively magnetic field strength) is measured, gyrochronology (respectively magnetochronology) analysis can be applied.} However, as long as the star departs from an isolated state, i.e., if a massive planet ($\rm M_p > 1~M_{jup}$) orbits at a close separation (less than 0.1 au) around the star, these techniques can no longer be used since star-planet tidal interaction could have modified the evolution of the surface rotation rate along the system's evolution \citep{Gallet18}. Consequently, the ages of numerous star-planet systems are currently not known and might appear younger than they really are. In this article we aim to provide the community with a reliable region where these empirical techniques can be used with confidence. We also present {the proof of concept of a} new age determination technique based on  modeling  the evolution of the star-planet tidal interaction, and on observation of the surface rotation rate of the host star and current location of the massive planet orbiting it.

\section{Star-planet tidal interaction model}

{In this section we briefly describe the numerical model used in this work as well as the explored grid of initial planetary systems.}

\subsection{Model}

In this work we combined the stellar rotational evolution model described in \citet{GB15} with the modified orbital evolution model used in \citet[][]{Bolmont16} \citep[see][for  details]{Gallet18}. The link between the two is done through the tidal dissipation induced by the gravitational interaction between the planet and the host star \citep[see][]{Gallet17b}. The stellar structure is provided by the stellar evolution code STAREVOL \citep[see][and references therein]{Amard15}.

The aim of this coupling is to follow the evolution of a given star-planet system, using a realistic approach. This code is specifically designed for stars between 0.3 (fully convective limit) and 1.2 $\rm M_{\odot}$ {(almost fully radiative limit)}, and for planetary systems composed of one close-in planet. {The tidal dissipation formalism used in our model is based on the parametrization of \citet{Bolmont16}, which  follows the simplified model by \citet{Ogilvie13}.}
{In this work, the star is assumed to be composed of a radiative core surrounded by a convective envelope both treated as regions of uniform densities. Moreover, in this work we only consider the frequency-averaged tidal dissipation and neglect the frequency response of the tides.}
{Since these are strong assumptions, the derived frequency-averaged tidal dissipation can be in error by about one or two orders of magnitude  compared to the actual tidal dissipation \citep{Ogilvie13}. In  Appendix \ref{tidal_intensity} we explore the impact on the results presented in this paper of decreasing or increasing the dissipation by two orders of magnitude.}
{We show that it does not affect  our conclusion qualitatively. The model also does not include the dissipation in the planet (which is still hardly theoretically constrained) or the magnetic star-planet interaction \citep[see][]{Strugarek17}. Moreover, the dissipation inside  the radiative core is also currently not physically described and hence not numerically included.}
{However, for hot Jupiters and in the case of circular orbit the typical evolutionary  timescale of planetary rotation rate and inclination evolution is so short ($10^5$ yr) that it can be safe to neglect the effects of tidal dissipation inside the planet \citep{Leconte10,Damiani18}.} 
{Taking into account  these extra dissipations could shorten the migration timescale during the MS phase and increase the departure from periods derived from wind-driven angular momentum loss, as supposed in gyrochronology analysis.}

{Finally, in this work we adopted an updated version of the relations from \citet{Gallet18} that link the initial stellar conditions to the stellar mass $\rm M_{\star}$ and initial rotational period $\rm P_{rot,init.}$:}
\begin{eqnarray}
\label{equaGB151}
\tau_{\rm c-e} &\approx& 8.01 \times \rm P_{\rm{rot,init}}^{0.69} \times \left(\displaystyle \frac{M_{\star}}{M_{\odot}}\right)^{-3.83} \rm{Myr},  \\
\label{equaGB152}
\tau_{\rm disk} &\approx& 2.24 \times \rm P_{\rm{rot,init}}^{0.61}  \times \left(\displaystyle \frac{M_{\star}}{M_{\odot}}\right)^{2.24} \rm{Myr},  \\
\label{equaGB153}
\rm K_1 &\approx& 41.2 + 37.6\left(\rm \displaystyle \frac{M_{\star}}{M_{\odot}}\right)^2-77.3\displaystyle \frac{\rm M_{\star}}{\rm M_{\odot}}.
\end{eqnarray}
{Here $\tau_{\rm c-e}$ is the coupling timescale between the radiative core and the convective envelope, $\tau_{\rm disk}$ the disk lifetime, and $\rm K_1$ the efficiency of the extraction of angular momentum by the stellar winds \citep[see][]{GB13,GB15,Gallet18}.} 

{Even if the detailed description of these quantities can be found in \citet{GB15}, we recall the meaning of each  of them. The coupling timescale $\tau_{\rm c-e}$ controls the characteristic timescale of internal angular momentum transport process. Short timescales correspond to strong coupling. We adjust these values so as to reproduce the rotational distribution of the early MS clusters (e.g., Pleiades, M50, and M35). The disk lifetime $\tau_{\rm disk}$ is the duration during which the surface rotation rate of the star is maintained constant \citep[see][]{Rebull04} due to the star-disk magnetic interaction. The disk lifetime is fixed by the observation of the rotational distribution of early pre-main-sequence (PMS) clusters (e.g., Orion, NGC 6530, and NGC 2362). Finally, $\rm K_1$ is the wind braking efficiency that was introduced in the \citet{Matt12} braking law. It controls the amount of angular momentum that is extracted by the stellar wind.}

\subsection{Grid of initial planetary systems}

{Using the PROBE\footnote{\textit{PeRiod and OrBital Evolution code}}} code described in \citet{Gallet18}, we computed a grid composed of  0.5, 0.7, and 1.0 $M_{\odot}$  stars with initial rotational periods between 1 and 11 days \citep[with $\Delta \rm P_{rot,init}$ = 0.2 days, and using a parametrization that follows][see above]{GB15,Gallet18}. We considered  2 $\rm M_{jup}$ and  5 $\rm M_{jup}$  planets initially located at 0.1, 0.3, 0.4, 0.45, 0.5, 0.55, 0.7, 0.8, 0.92, 0.94, 0.96, 0.98, and 1.0 $\rm R_{co}$, with the corotation radius $\rm R_{co}$ given by
\begin{equation}
\label{corot}
\rm R_{\rm co} = \left( \displaystyle \frac{\G M_{\star} }{\Omega_{\star}^2}  \right)^{1/3} = \left( \displaystyle \frac{\G M_{\star} P_{\rm rot,\star}^2 }{(2\pi)^2}  \right)^{1/3},
\end{equation}
where $\rm P_{\rm rot,\star}$ is the surface rotation period of the host star, $\G$ the gravitational constant, $\rm M_{\star}$ the stellar mass, and $\Omega_{\star} = 2\pi / \rm P_{\rm rot,\star}$ the surface angular velocity of the host star.

{In our model, we assume as initial conditions the rotation period $\rm P_{init}$ and the semi-major axis (SMA) at the time $\rm t_0$. The age $\rm t_0$ is the starting age of our simulations, and corresponds to the age at which the disk dynamically and magnetically decouples from the star. It corresponds here to the disk lifetime (i.e., the end of the disk locking phase) that occurs during the early PMS phase. }

{We also  note that we only consider planets initially inside of the co-rotation radius. This is motivated by the fact that it is well documented \citep{Laine08,Laine12,Chang12,Bolmont12,Bolmont16} that planets  initially (i.e., at the disk lifetime) outside of the corotation radius tend to migrate outward, while the central star is contracting and thus accelerating toward the zero age main sequence. As a consequence, these planets will move outward, while the corotation radius moves inward. The direct effect of this opposite evolution is to drastically reduce the action of the tides induced by the planet on the stellar surface rotation rate. In this case, the central star will simply follow the same angular velocity evolution as its counterpart without a planet, which allows the use of the surface rotation rate, for example via gyrochronology (see Sect. \ref{gyro_appli}), with fairly high confidence. }

\section{Gyrochronology and its domain of applicability}

{By using the numerical model and method presented in the previous section, we investigate the limit of the application of gyrochronology in the case of massive close-in planetary systems.}

\label{gyro_appli}
\begin{figure*}[!ht]
    \begin{center}
            \includegraphics[width=1.0\linewidth]{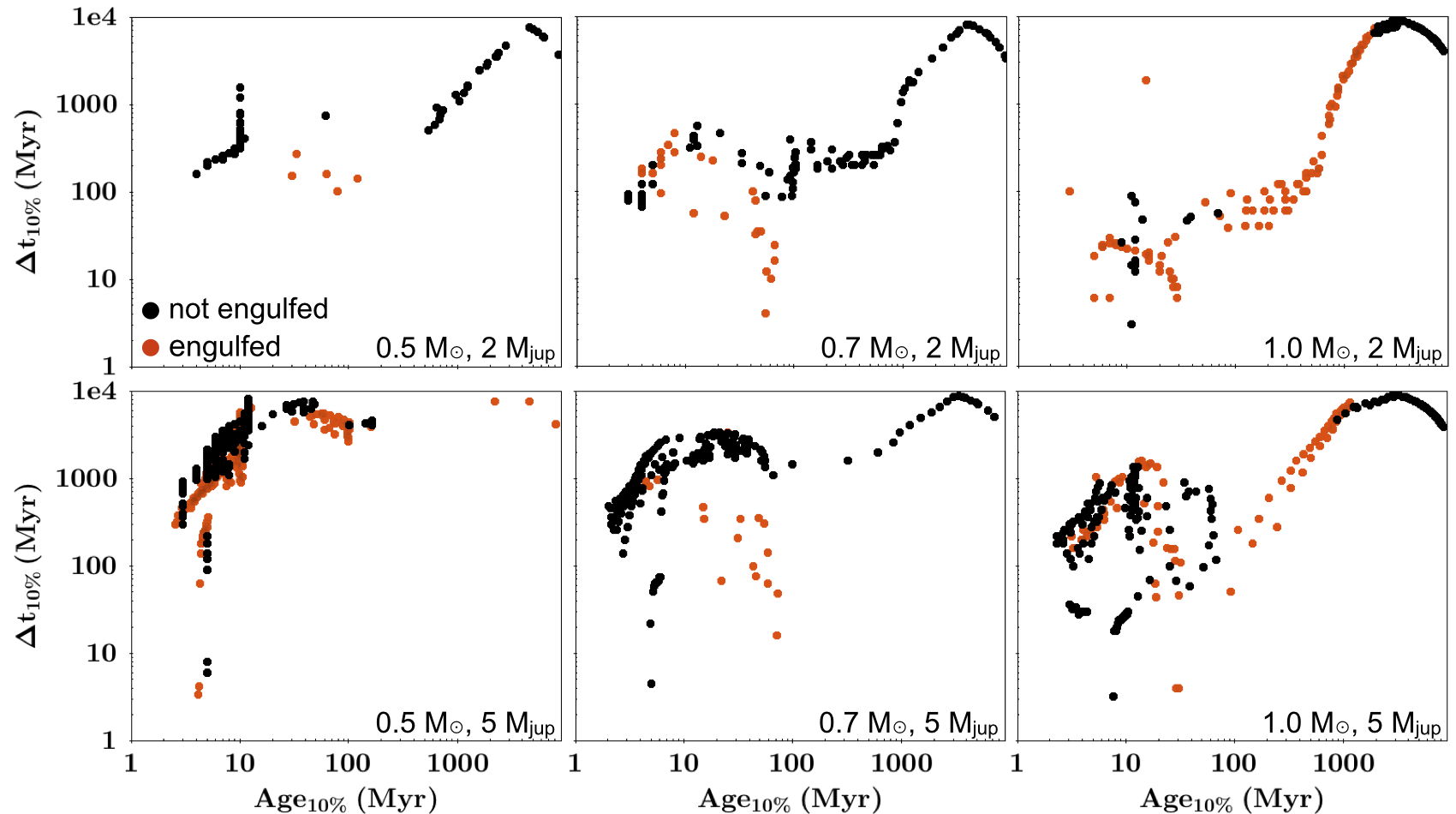}  
        \caption{Duration for which the surface rotation rate of a star  in a star-planet system  departs by more than 10\% compared to the same but isolated star (see sect. \ref{nomanclature} for the definition of the variables). The black dots correspond to  planets still orbiting the star, while the red dots are planets that have been engulfed.}
        \label{Deltat}%
    \end{center}
\end{figure*}

\subsection{{Nomenclature and definition of the variables}}
\label{nomanclature}
Gyrochronology was initially proposed by \citet{Barnes03} as a technique to estimate the age of isolated stars. It is based on the behavior  of the surface rotation rate of the stars between 0.3 and 1.0 $\rm M_{\odot}$ that seems to evolve as $\rm t^{-1/2}$ during the MS phase \citep[see][]{Sku72}. 

{In this section, and for the sake of clarity, we  provide  the definitions of numerous variables that will be used in the following:}
\begin{itemize}
    \item $\rm t_0$ is the starting age of our simulations. {In our model it corresponds to the end of the star-disk magnetic interaction phase ($\rm \tau_{disk}$) that depends on the initial properties of the star}.
    \item $\rm P_{init}$ is the stellar rotation period at $\rm t_0$.
    \item $\rm SMA_{init}$ is the planetary SMA at $\rm t_0$.
    \item $\rm P_{rot,isol.}$ is the rotation period of a star without a planet.
    \item $\rm P_{rot,\star}$ is the rotation period of a  star  with an orbiting planet.
    \item $\rm \delta P_{rot} = 1 - P_{rot,isol.}/P_{rot,\star}$ is the {relative} rotational departure between a star with  and without planet.
    \item $\rm Age_{x\%}$ is the age at which the {relative} rotational departure $\rm \delta P_{rot}$ becomes greater than $\rm x\%$ (because of a planetary engulfment or strong tidal interactions).
    \item $\rm P_{rot,x\%}$ is the rotation period of a star at the specific age $\rm Age_{x\%}$.
    \item $\rm P_{orb,x\%}$ is the orbital period of a planet at the specific age $\rm Age_{x\%}$.
    \item $\rm P_{rot,isol.,x\%}$ is the rotation period of a  star but without a planet at the specific age $\rm Age_{x\%}$. It is used as a comparison value to quantitatively extract the impact of the presence of a massive planet on the surface rotation rate of the stars. 
    \item $\rm \Delta t_{x\%}$ is the time during which the {relative} rotational departure $\rm \delta P_{rot} \ge x\%$. For example, $\rm \Delta t_{0\%}$ corresponds to the time it takes for a star to return to the rotational behavior of an isolated star. 
    \item $\rm Age_{gyro}$ is the age estimated using a gyrochronology tool and based on $\rm P_{rot,x\%}$. It corresponds to the age  estimated using the observed stellar rotation period without information about the rotational history. 
    \item $\rm Age_{gyro, isol.}$ is  the ``real'' age of a star estimated using  gyrochronology and based on $\rm P_{rot,isol.,x\%}$. It corresponds to the true gyrochronologic age of a star that has never experienced an  impact of a planetary companion. 
\end{itemize}
We note that $\rm Age_{gyro, isol.}$ and $\rm Age_{gyro}$ are semi-synthetic ages since they are estimated using the rotation rate provided by the numerical model and the gyrochronology tools calibrated on the observations. The difference between $\rm Age_{gyro, isol.}$ and $\rm Age_{gyro}$ highlights the possible error on the age estimation that can be introduced by the presence of a massive close-in planet.
Moreover, in the case of a perfect age-$\rm P_{rot,\star}$ relationship, $\rm Age_{gyro, isol.}$ and $\rm Age_{x\%}$ should, by definition, be identical. However, since $\rm Age_{x\%}$ is directly obtained from the stellar model STAREVOL, while $\rm Age_{gyro}$ is calibrated using observations, the  two ages are different (see Table \ref{tab1}).

{In this article we used the gyrochronology calibration described in \citet{Delorme11} that is slightly different from that proposed in \citet{Barnes03,Barnes07,Barnes10}. In \citet{Delorme11} the calibration of the gyrochronology is done using the SuperWASP cluster  data \citep{WASP} in addition to the properties of the Hyades cluster} 
\begin{eqnarray}
\rm Age_{gyro} = Age_{Hyades}*(P_{rot,obs.}/P_{Hyades})^{1/0.56}~Myr,
\end{eqnarray}
{where the period-color relation in the Hyades is given by}
\begin{eqnarray}
\rm P_{Hyades}= 10.603+ 12.314*(J-K_{s}-0.57)~days,
\end{eqnarray}
{with $\rm Age_{Hyades}=625~Myr$. The (J-$\rm K_s$)-mass relation is given by the YREC isochrones extracted at 600 Myr for solar metallicity stars \citep[Z = 0.01757, see][]{YREC}\footnote{http://www.astronomy.ohio-state.edu/iso/empirical.html}.}

{Finally, gyrochronology analysis can only be applied to systems whose {real} age is older than about 100-200 Myr for a 1.0 $\rm M_{\odot}$ star and 500-600 Myr for a 0.6 $\rm M_{\odot}$ star \citep{Barnes10,Delorme11}.}

\subsection{Impact of tidal interaction on rotation}

Figure \ref{DeltaP3D} shows the evolution of $\rm P_{rot,\star}$(t) as a function of the SMA(t) expressed in orbital period $\rm P_{orb}(t)$, for the three stellar masses (0.5, 0.7, and 1.0 $M_{\odot}$), and the two planetary masses (2 $\rm M_{jup}$ and 5 $\rm M_{jup}$) considered in this work. The relation between SMA and $\rm P_{orb}$ is given by
\begin{eqnarray}
\rm P_{orb} = 2\pi \sqrt{\frac{\rm SMA^3}{\rm \mathcal{G}(M_{\star}+M_p)}}.
\end{eqnarray}
The color gradient corresponds to the variation in $\delta \rm P_{rot}$ for each pair $[\rm P_{rot,\star}(t)-SMA(t)-M_{\star}-M_p]$.

Figure \ref{DeltaP3D} displays the domain of validity of the gyrochronology analysis. In this figure the brightest part corresponds to the region where the star is 10\% faster ($|\delta \rm P_{rot}| \geq 10\%$) than an isolated star, which corresponds to an error of about 20\% on the age estimation using the gyrochronology analysis \citep[$\rm \Omega_{\star} \propto t^{-\alpha}$, with $\alpha$ around 0.50-0.56, see][]{Barnes07,Mamajek08,Delorme11,Angus15}. This region, which depends on the stellar and planetary mass, is located in the left part of each of these plots. For the {1.0} $\rm M_{\odot}$ star, this region is positioned around $\rm P_{orb}(t)< 4.2$ days (SMA(t) $\lesssim$ 0.05 au) and $\rm P_{rot}(t)$ between 10 and 50 days. Increasing the planetary mass slightly extends this region toward higher $\rm P_{orb}(t)$ (SMA(t)).


Figure \ref{Deltat} shows the characteristic time during which the impact of the star-planet tidal interaction on the surface rotation rate of the star remains above $\rm \delta P_{rot}$ = 10\%, {which corresponds to the typical systematical error of the gyrochronology analysis.} If the departure $\rm \delta P_{rot}$ is above 10\% during the PMS (age $\lesssim$ 100 Myr) then the characteristic timescale $\rm \displaystyle \Delta t_{10\%}$ decreases with increasing stellar mass. {For all the stellar masses,} the duration $\rm \displaystyle \Delta t_{10\%}$ {ranges between} a few tens or hundreds of Myr and several Gyr. However, the effect of the impact of the star-planet tidal interaction is visible at longer ages when the planet is engulfed during the MS phase (due to the stable internal structure of the stars during that phase).

If a planet more massive than about 1 $\rm M_{jup}$ and located below 0.1 au is detected orbiting its host star, then applying the gyrochronology would lead to a systematic underestimation of the age of the system by at least 20\%. If no planets are detected, but  there is a suspicion of planetary engulfment, then the gyrochronology analysis could also lead to significant systematics errors. 

In some cases, the gyrochronology cannot be applied even several Gyr after the engulfment (see Fig. \ref{Deltat}). The GJ 504 system \citep{Fuhrmann15,D'Orazi17,Bonnefoy18} is a striking example of the difficulty of determining the age of a given system even with the use of multiple age estimation techniques. In this case, the available techniques in the literature provide colliding results leading to either a young system (age less than 150 Myr) or an advanced system (age above 2 Gyr).

{In the case of GJ 504, the error on the estimation of the age is clearly beyond the 20\% error limits that we considered in the previous sections. To investigate whether the presence of a massive planet around a star can produce  such a discrepancy in the age estimation, we also considered very strong star-planet interactions with $\rm \delta P_{rot} \ge 0.9$.}

{Table \ref{tab1} shows the difference in the age estimation using a gyrochronology tool \citep[as used in][]{Delorme11} between an isolated star and the same star but with a massive planet as companion. We extracted the rotation period $\rm P_{rot,90\%}$, the orbital period $\rm P_{orb,90\%}$ (when the planet is still orbiting the star), and the age $\rm Age_{90\%}$ at which the rotation of the star departs by more than 90\%  compared to the same but isolated star. 
By using $\rm P_{rot,90\%}$ and $\rm P_{rot,isol.,90\%}$, we estimated the gyrochronology ages ($\rm Age_{gyro}$ and $\rm Age_{gyro, isol.}$) of these systems so as to extract the error on the age estimation imputable to the star-planet interaction. The difference in age can reach  several Gyr. While a systematic error of more than 90\% on $\rm \delta P_{rot}$ only lasts a few tens or hundreds of Myr (see Table \ref{tab1}, $\rm \Delta t_{90\%}$), in the case of a 10\% error $\rm \Delta t_{10\%}$ can reach a few Gyr (see Fig. \ref{Deltat}). In Table \ref{tab1}, we list the data for the 2 $\rm M_{jup}$ and 5 $\rm M_{jup}$  planets {(for $\rm \delta P_{rot} \ge$ 0.1, 0.25, 0.5, 0.75, see  Appendix \ref{explodeltap})}.}  

{Recently, \citet{Metcalfe18} investigated the limit of  gyrochronology from a more intrinsic point of view. In their paper they highlighted a possible breakdown of the gyrochronology relations for stars in their mid-MS phase due to the shutdown of their dynamo driven spin-down process. They showed that during this advanced phase, chromospheric activity should provide a more reliable age estimation than gyrochonology analysis.}

\begin{table*}
\caption{Error on the estimation of the age provided by the gyrochronology analysis for stars that have experienced a {relative} rotational departure greater than $|\rm \delta P| = 0.9$. These data are  for a 2--5 $\rm M_{jup}$  planet. The long dash  ``--''  means that the planet is already engulfed at $\rm Age_{90\%}$. {Only part of the full dataset is shown; the rest is available on demand.} See Sect. \ref{nomanclature} for the definition of each of these quantities. {The quantities written in italics indicate {ages that are estimated outside of the application range of gyrochronology.}}}
\label{tab1}
\begin{tabular}{|c|c|c|c|c|c|c|c|c|c|}
\hline
Mass & $\rm P_{init}$ & $\rm SMA_{init}$ &  $\rm Age_{90\%}$ & $\Delta t_{90\%}$  & $\rm P_{rot,90\%}$ & $\rm P_{orb,90\%}$ & $\rm P_{rot,isol.,90\%}$ & $\rm Age_{gyro}$ & $\rm Age_{gyro, isol.}$ \\
($M_{\odot}$) & (days) & (au) & (Myr) & (Myr) & (days) & (days) & (days) & (Myr) & (Myr) \\ 
\hline
\hline
0.5      &      4.60     &      0.0236   &      782/74   &      100/268  &       6.40/0.85        &      --/0.1647        &      13.24/1.87       &       159/\textit{4}   &      583/\small\textit{18} \\
0.5      &      5.00     &      0.0250   &      722/66   &      100/296  &       6.79/0.86        &      --/0.1641        &      13.31/2.03       &       177/\textit{4}   &      588/\small\textit{21} \\
0.5      &      5.40     &      0.0239   &      1052/98  &      150/284  &       7.81/1.05        &      --/0.1653        &      16.13/2.51       &       227/\textit{6}   &      830/\small\textit{30} \\
0.5      &      6.00     &      0.0231   &      2002/102         &      250/320  &       11.14/1.19       &      --/0.1666        &      22.85/2.78       &       428/\textit{8}   &      1545/\small\textit{36} \\
0.5      &      8.60     &      0.0196   &      6203/10  &      750/173  &       14.64/2.80       &      0.2642/0.1650    &      41.32/6.31       &       697/\textit{36}  &      4450/\small\textit{155} \\
\hline
\hline
0.7      &      5.80     &      0.0224   &      763/66   &      41/78    &       2.79/2.06        &      --/0.1610        &      13.03/4.68       &       \textit{42}/\textit{24}  &      650/105 \\
0.7      &      7.40     &      0.0198   &      622/7    &      42/56    &       2.78/2.90        &      --/--    &      12.87/6.08       &      \textit{41}/\textit{44}  &       636/167 \\
0.7      &      7.60     &      0.0202   &      985/8    &      70/57    &       7.45/2.68        &      --/--    &      16.07/5.88       &      239/\textit{39}  &       946/157 \\
0.7      &      7.80     &      0.0205   &      1355/11  &      100/56   &       8.33/2.32        &      --/--    &      18.92/5.38       &      292/\textit{30}  &       1266/134 \\
0.7      &      8.00     &      0.0209   &      1755/21  &      100/50   &       9.51/1.48        &      --/--    &      21.91/4.03       &      370/\textit{13}  &       1645/\textit{80} \\
0.7      &      8.20     &      0.0212   &      2155/165         &      150/100  &       11.94/4.80       &      --/--    &      24.71/10.41      &      557/109  &       2039/435 \\
0.7      &      8.80     &      0.0222   &      3456/946         &      400/160  &       16.90/8.09       &      --/--    &      33.09/16.37      &      1034/278         &       3435/978 \\
0.7      &      9.00     &      0.0226   &      3906/1206        &      550/200  &       17.95/8.78       &      --/--    &      35.08/18.34      &      1152/321         &       3812/1198 \\
0.7      &      10.00    &      0.0242   &      7156/2656        &      1100/600         &       18.99/13.37      &      0.2643/0.1607    &      47.04/27.76      &       1274/681         &      6437/2511 \\
0.7      &      10.40    &      0.0248   &      8707/3407        &      1500/900         &       19.48/14.35      &      0.2642/0.1575    &      50.55/32.55      &       1333/773         &      7319/3335 \\
\hline
\hline
1.0      &      4.00     &      0.0222   &      491/33   &      14/16    &       1.70/1.92        &      --/--    &      7.00/3.81        &      \textit{38}/\textit{48}  &       482/162 \\
1.0      &      5.00     &      0.0258   &      2256/22  &      100/12   &       8.26/1.01        &      --/--    &      17.78/2.98       &      647/\textit{15}  &       2544/105 \\
1.0      &      5.00     &      0.0286   &      1506/102         &      50/24    &       7.28/2.31        &      --/--    &      14.08/4.51       &      516/\textit{66}  &       1676/220 \\
1.0      &      6.00     &      0.0194   &      178/14   &      9/15     &       1.65/2.19        &      --/--    &      5.59/4.26        &      \textit{36}/\textit{61}  &       322/198 \\
1.0      &      6.00     &      0.0323   &      5407/21  &      800/12   &       13.96/0.96       &      --/--    &      26.81/3.79       &      1652/\textit{14}         &       5295/161 \\
1.0      &      7.00     &      0.0322   &      6808/4208        &      1150/1500        &       15.91/12.37      &      --/--    &      30.30/23.79      &      2085/1330        &       6588/4275 \\
1.0      &      7.00     &      0.0215   &      706/186  &      22/22    &       1.71/0.34        &      --/--    &      9.21/6.09        &      \textit{39}/\textit{2}   &       785/375 \\
1.0      &      8.00     &      0.0313   &      6309/3759        &      900/1150         &       15.07/11.56      &      --/--    &      28.97/22.27      &      1892/1179        &       6078/3800 \\
1.0      &      8.00     &      0.0235   &      1459/569         &      50/40    &       6.97/4.15        &      --/--    &      13.68/8.67       &      477/189  &       1591/705 \\
1.0      &      9.00     &      0.0339   &      7710/5060        &      1150/1800        &       17.01/13.47      &      --/--    &      32.78/25.74      &      2348/1550        &       7580/4922 \\
1.0      &      9.00     &      0.0254   &      2460/1160        &      150/50   &       9.38/4.78        &      --/--    &      17.93/12.08      &      812/243  &       2581/1274 \\
1.0      &      10.00    &      0.0363   &      8811/6261        &      1300/2200        &       18.97/14.91      &      --/--    &      36.37/28.57      &      2856/1857        &       9129/5931 \\
1.0      &      11.00    &      0.0290   &      4912/2662        &      400/400  &       12.64/9.34       &      --/--    &      24.91/18.37      &      1383/805         &       4642/2694 \\

\hline 
\end{tabular}
\end{table*}

\section{Tidal-chronology}

Following the work of \citet{Gallet18} we developed a new age estimation technique based on the measurement of  the surface rotation rate of the star and the location of the planet around it. It relies on the fact that the star-planet interaction produces {a rotation cycle that can be used to estimate the age of a given close-in system. The four steps of this cycle are as follows}:
\begin{itemize}
\item 1) The system starts its evolution with an initial condition ($\rm P_{rot,init}$-$\rm SMA_{init}$); 
\item 2) Given the value of the initial rotation rate of the star and its internal structure at the initial time $t_0$, the efficiency <$\mathcal{D}$>$_{\omega}$ of the dissipation of the star is estimated;  
\item 3) The value <$\mathcal{D}$>$_{\omega}$ of the dissipation determines the orbital evolution of the planetary companion;
\item 4) The planetary orbital evolution then modifies the surface rotation rate of the star. 
\end{itemize}
The star-planet system starts again at the first step with a new pair $\rm P_{rot}$-$\rm SMA$.

Since the rate of the evolution of the SMA strongly depends on its value \citep[see Eqs. 3-6 in][]{Gallet18}, a given observed pair  $\rm P_{\rm rot,\star}(\rm t)$-$\rm{SMA(t)}$ is thus only produced for a small range of possible ages and initial conditions, {as long as the observed rotational period of the star is longer than 10 days} (see Sect. \ref{degeneracies}). {As in gyrochronology, for a rotational period below 10 days, which corresponds to a system younger than about 100 Myr, the tidal-chronology technique will provide degenerated solutions.} 
\begin{figure*}[!ht]
    \begin{center}
        \includegraphics[width=\linewidth]{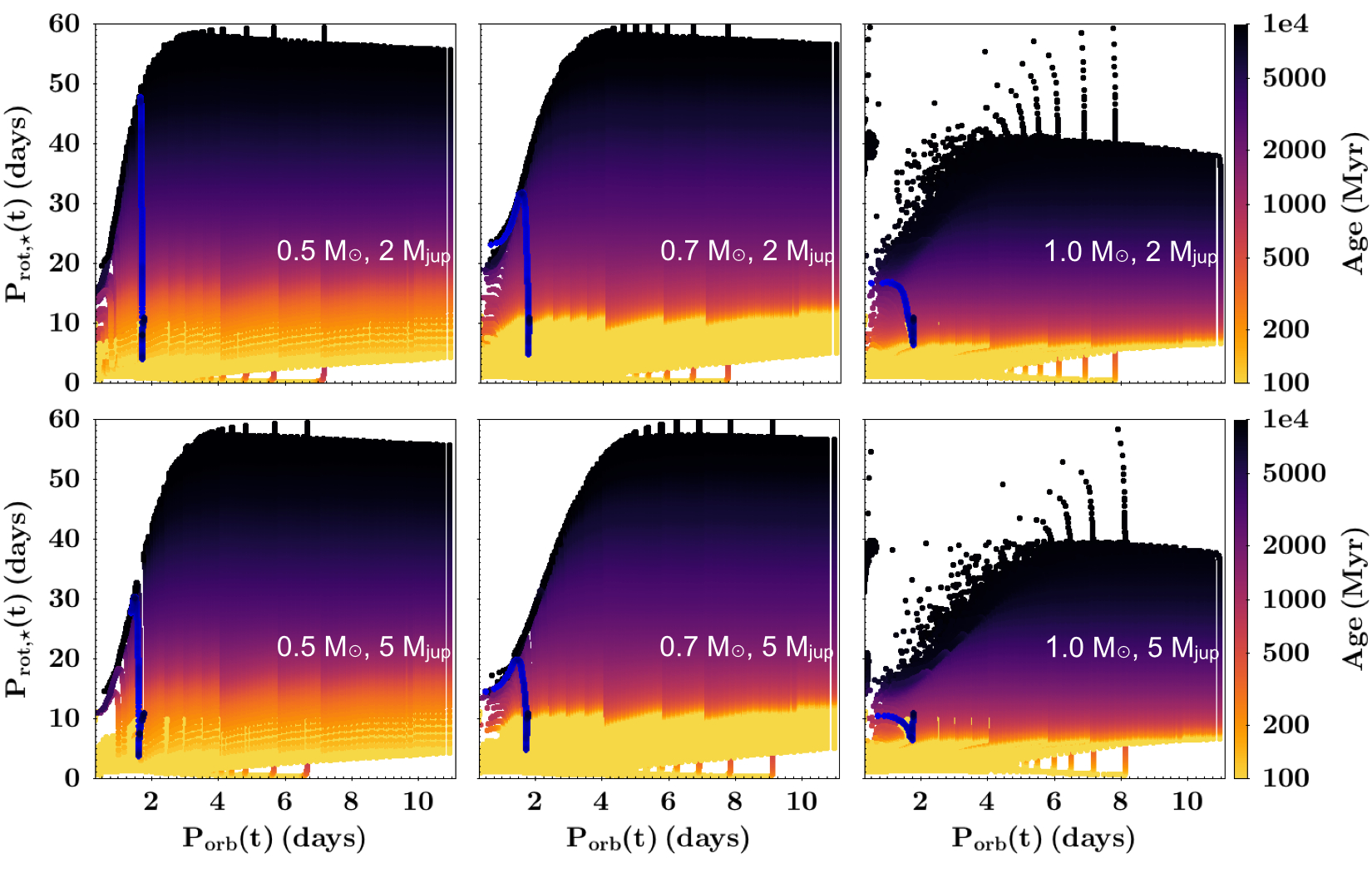}
        \caption{Synthetic $\rm P_{\rm rot,\star}$(t) and $\rm P_{orb}(t)$ estimated for a system composed of a 0.5, 0.7, and 1.0 $M_{\odot}$ star around which orbits a 2 $M_{\rm jup}$ planet (upper panel) and a 5 $M_{\rm jup}$ planet (lower panel). $\rm P_{rot,init}$ = 1-11 days (with $\Delta \rm P_{rot,init}$ = 0.2 days) and $\rm SMA_{init}$ is between 0.1 and 1.0 $\rm R_{co}$ \citep{Gallet18}. The color gradient indicates the age (in Myr) at which the pair $\rm P_{\rm rot,\star}(t)-P_{orb}(t)$ is extracted. Only planetary systems where the planet is still orbiting the star are plotted. {The blue lines (composed of plain circles) depict the evolution of a given star-planet system ($\rm P_{rot,init}=10.8$ days and $\rm SMA_{init}= 30\% R_{co}$) in the $\rm P_{rot,\star} (t)-\rm P_{orb} (t)$ plane.}}
\label{Degemass3D}%
    \end{center}
\end{figure*}

\subsection{Description {of the tidal-chronology technique}}

To determine the age of a planetary system we thus computed a grid of star-planet system evolutions composed of a star with $\rm P_{\rm rot, init}$ between 1 and 11 days and a planet's initial SMA between 0.1 and 1.0 $R_{\rm co}$, for the given stellar and planetary mass that match the observed system's properties. 

We chose to express the initial SMA as a function of $R_{\rm co}$, rather than fixing the range of initial planetary orbits, so as to ensure a good coverage of the initial conditions for which the planet orbits inward ($\rm{SMA}_{\rm init} \leq R_{\rm co}$,  $1.8\times10^{-3} \rm < SMA_{init} (au) < 0.09$). {In the case of outward migration, the surface rotation rate of the star is less impacted by the star-planet tidal interaction.}

We then explored the grid so as to extract  age at which the observed pair $\rm P_{\rm rot,obs}$-$\rm SMA_{obs}$ is retrieved. The resulting grid is composed of 714 rotational-orbital evolutions (51 initial rotational periods $\times$ 14 initial SMAs) for each star-planet system. We finally estimate the departure of the observed pair $\rm P_{\rm rot,obs}$-$\rm SMA_{obs}$ to each of the $\rm P_{\rm rot,\star}$-$\rm SMA_\star$ pairs from the grid using the  expression 
\begin{eqnarray}
{S^2 = \frac{(\rm SMA_{\rm \star}-\rm SMA_{\rm obs})^2}{ \rm \sigma_{\rm SMA_{obs}}^2} + \frac{(\rm P_{\rm orb,\star}-\rm P_{\rm orb,obs})^2}{ \rm \sigma_{\rm P_{orb,obs}}^2},}
\end{eqnarray}
{where $\displaystyle \sigma_{\rm SMA_{obs}}$ and $\displaystyle \sigma_{\rm P_{orb,obs}}$ are the standard deviation of the observed $\rm SMA_{obs}$ and $\rm P_{orb,obs}$.}

We also applied a 3D interpolation method using the \textit{Python-SciPy griddata} routine \citep[to interpolate unstructured 3D data, see][]{Python}.

\subsection{Limits, degeneracies, and framework of the tidal-chronology}
\label{degeneracies}
To investigate the degeneracies of this technique we considered a 0.5, 0.7, and 1.0 $M_{\odot}$  stars and a 2 and 5 $M_{\rm jup}$  planets  and computed a grid composed of $\rm P_{rot,init}$= 1-11 days (with $\Delta \rm P_{rot,init}$ = 0.2 days) and $\rm SMA_{init}=$ 0.1, 0.3, 0.4, 0.45, 0.5, 0.55, 0.7, 0.8, 0.92, 0.94, 0.96, 0.98, and 1.0  $\rm R_{co}$ \citep[see][]{Gallet18}. 

Figure \ref{Degemass3D} shows $\rm P_{\rm rot,\star}$(t) as a function of $\rm P_{orb}$(t), $M_{\star}$, and $\rm M_p$, and displays the age as a color gradient. {In this figure, $\rm P_{orb}$(t) is between 0.5 and 12 days and $\rm P_{rot,\star}$(t) between 0 and 50 days}. It shows that the solutions are indeed degenerated when using only $\rm P_{\rm rot,\star}$ or SMA, but that they are lifted when using both quantities simultaneously. Additionally, the information whether the planet is still  orbiting the star adds another criterion and helps {remove} these degeneracies. 

\begin{figure}[!ht]
\begin{center}
   \includegraphics[width=0.9\linewidth]{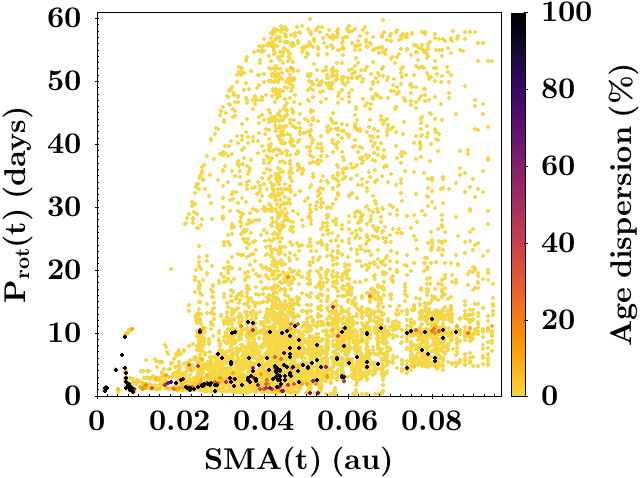} 
   \caption{{Randomly selected $\rm P_{\rm rot,\star}$(t)-$\rm SMA$(t) pairs (see text) with their associated age dispersion $\rm (t_{max}-t_{min})/t_{min}$.}}
\label{Count}%
\end{center}
\end{figure}

{Figure \ref{Degemass3D} shows that the age linearly increases for increasing $\rm P_{\rm rot,\star}$(t) and highlights that for $\rm P_{\rm rot} (\rm t) \gtrsim 5$ days the degeneracy in the estimation of the age is lifted when using the information about SMA(t). 
{With Fig. \ref{Count} we can see that the age dispersion is below 10\% for most of the $\rm P_{\rm rot,\star}$-$\rm SMA$ space. The degeneracy of the technique is {thus} quite low and might be a concern only for young systems with ages less than about 100 Myr (which are also beyond the traditional gyrochronology application range) that globally correspond to systems in which $\rm P_{\rm rot} (\rm t)$ $\lesssim$ 10 days.} The degeneracy is mainly due to the high temporal resolution of the models during the PMS that display almost identical pairs $\rm P_{\rm rot} (\rm t)-SMA(t)$ for slightly different ages.}


The evolution of the star-planet systems in Fig. \ref{Degemass3D} starts from the bottom of the $\rm P_{\rm rot,\star}$(t)-$\rm P_{orb}$(t) space, between 1.8$\times 10^{-3}$ au ($\rm P_{\rm rot,init}$ = 1 day, $\rm SMA_{init}$ = 0.1 $\rm R_{co}$) and 0.09 au ($\rm P_{\rm rot,init}$ = 11 days, $\rm SMA_{init}$ = $\rm R_{co}$) and  $\rm P_{\rm rot,init}$ = 1-11 days. Without planetary migration, the evolution of a star-planet system is depicted by a straight line transiting from the brightest (age $\approx$ 100 Myr) to the darkest (age > 1 Gyr) parts of the figure. It concerns star-planet systems in which the planet is initially located around $\rm R_{co}$. 

{Figure \ref{Count} shows the age dispersion for given $\rm P_{\rm rot,\star}$(t)-$\rm SMA$(t) pairs. It highlights the small degeneracy of the tidal-chronology technique when determining the age of a given planetary system from its observed $\rm P_{\rm rot,\star}$(t)-$\rm SMA$(t) pair. In Appendix \ref{precdeg} we explore the impact of the precision of $\rm P_{\rm rot,\star}$ on the degeneracy of the technique.} 

{To produce this figure we first bin the $\rm P_{\rm rot,\star}$-$\rm SMA$ space using the classical errors on $\rm P_{\rm rot,\star}$ and $\rm SMA$: $\Delta \rm P_{\rm rot,\star} =0.001$ days \citep{K2} and $\Delta \rm SMA =0.00018$ au \citep{Hellier11}. Given the range of $\rm P_{\rm rot,\star}$ and $\rm SMA$ of our models, this binning produced an image composed of 30 million pixels.}

{To allow a fast exploration of the parameter space, we then randomly selected 10000 pairs and performed a loop over the whole dataset of models so as to look at the age range, i.e., the minimum and maximum age $\rm t_{min}$ and $\rm t_{max}$, associated with each of these random pairs.} {We finally estimated the age dispersion $(\rm t_{max}-t_{min})/\rm t_{min}$ associated with each of these pairs and plot them in Fig. \ref{Count}.}

In the case of migrating planets, the evolution of star-planet systems in the $\rm P_{\rm rot,\star}$(t)-$\rm P_{orb}$(t) space depends on their initial conditions. As the star evolves, its surface rotation rate is impacted by the inward migration of the planet. The star-planet system thus moves toward  smaller SMA ($\rm P_{orb}$) and rotation period (if the acceleration torque produced by the planet is stronger than the braking torque of the stellar winds or the PMS contraction torque, see the blue tracks in Fig. \ref{DeltaP3D}). 

The hypothesis of this paper implies that the orbit of the planet is initially circular and coplanar, and remain so, with respect to the equator of the star during the whole stellar evolution. The results above and the tidal-chronology technique are hence only valid within this framework. {{Moreover}, since we simply present the proof of concept of such a technique, for now, the tidal-chronology should  be used with caution and in addition to other age determination tools. In the future, and with the increase in the completeness and complexity of our tidal dissipation modeling, we hope that this technique will be reliable enough to derive reliable stellar ages by itself.}

{In this paper, and because of the nature of the tidal dissipation formalism we use, we only consider isolated star-planet systems composed of one planet that orbits one star because the model that we use does not allow us to follow the evolution of multiplanetary systems. Hence, we do not consider Kozai-Lidov resonance \citep{Kozai,Lidov}, which  is present in n-body systems (with n > 3) for a given body with an {orbital} inclination greater than 39$^{\circ}$.} {Finally, we  also mention the work of \citet{Wu11} on secular chaos in the evolution of the properties of almost coplanar multiplanetary  systems. They pointed out that strong variations in the orbits of the planets can occur in these kind of system on timescales ranging from a few tens of Myr   to a few Gyr.}



\subsection{Observational case: WASP-43}


{To extract the order of magnitude of the difference in age estimation between standard gyrochronology and tidal-chronology, we decided to apply and illustrate our technique on a specific planetary system, namely WASP-43, for which the rotation rate of the star is well known, as is the location of the orbiting planet. }

WASP-43 is a K7V star (corresponding to 0.717 $\pm$ 0.025 $\rm M_{\odot}$), with $T_{\rm eff} = 4520 \pm 120$ K and solar metallicity [Fe/H] =  0.01 $\pm$ 0.012. Around this star orbits a 2.052 $\rm M_{\rm jup}$  planet \citep{Hellier11}. The rotation period of the star is estimated at 15.6 $\pm$ 0.4 days and the planet is observed at a distance of 0.01526 au from the star. {The standard deviations for WASP43 are $\displaystyle \sigma_{\rm SMA_{obs}} = 0.00018$ au and $\displaystyle \sigma_{\rm P_{orb,obs}} = 0.4$ days \citep{Hellier11}.} The system's eccentricity is very low {\citep[0.0035, see also][for other estimations of hot-Jupiter system parameters]{Bonomo17}} and the inclination is quite high (82 degree), which allows us to apply, {at the first order}, the dynamical tides formalism of our numerical code. 

It is not certain, however,  that the eccentricity was this low during the whole system evolution. The main effect of increasing the eccentricity of the systems on the evolution of the dynamical tide is the excitation of multiple frequencies, which can boost the tidal dissipation and thus the evolution of the SMA. These effects are expected to be the strongest during the PMS phase {(Emeline Bolmont, private communication)}. The same kind of effect can appear for the equilibrium tide. High eccentricity can lead to an increase in the transfer of angular momentum from the planetary orbit to the stellar rotation, which implies a faster rate of acceleration of the central star.
\begin{figure}[!ht]
\begin{center}
   \includegraphics[width=1.0\linewidth]{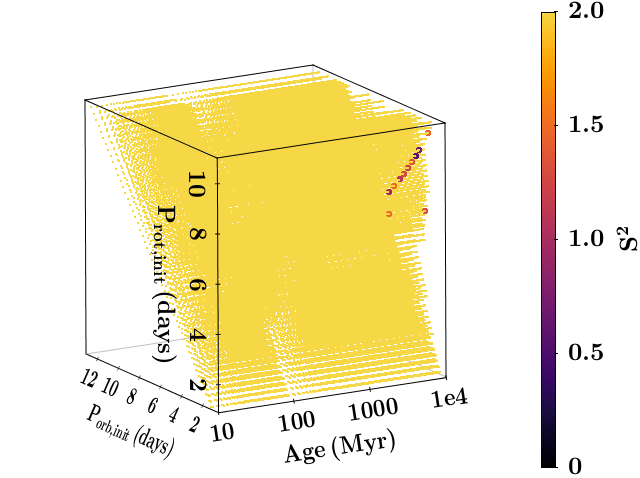} 
   \caption{Map of $\rm S^2$ for the WASP-43 system as a function of time, $\rm P_{\rm rot,init}$, and $\rm P_{orb,init}$. }
\label{3D}%
\end{center}
\end{figure}
Figure \ref{3D} shows (as a color gradient) the value of $\rm S^2$ as a function of time, $\rm{SMA}_{init}$, and $\rm P_{\rm rot,init}$. Using the gyrochronology technique based on the \citet{Barnes07} calibration, the age of WASP-43 was estimated at 400 Myr by \citet{Hellier11}. {With the more recent calibrations of \citet{Barnes10} and \citet{Delorme11}, a gyrochronologic age of 0.9-1.0 Gyr can be estimated for WASP43.} Using our tidal-chronology technique, the most probable solutions suggest an older system with an age between 2.4 and 9.2 Gyr (see Fig. \ref{3D}) {with a minimum $\rm S^2$ of 0.58 at 5.85 Gyr}. A 3D linear interpolation of the observed pair leads to an age estimation of 3.5${\pm 1.1}$ Gyr, which is consistent with our $S^2$ exploration. {The difference between the two estimated ages is also consistent with the global age shift that can be introduced by a massive planet (see Table \ref{tab1}).} 

\subsection{{Impact of the high uncertainty on the tidal dissipation intensity factor}}

{Since the efficiency of the tidal dissipation is currently not well  constrained in theory, we decided to explore to what extent the tidal-chronology technique depends upon this intensity. In this section, we thus investigate the impact of a 2 $\rm M_{jup}$  planet on the surface rotation rate of a 0.7 $\rm M_{\odot}$  star and consider a tidal dissipation intensity two orders of magnitude  higher (<$\rm {\mathcal{D}}$>$_{*100}$) or lower (<$\rm {\mathcal{D}}$>$_{/100}$) than  the averaged values (<$ \rm {\mathcal{D}}$>) from \citet{Gallet17}.}

{The results are summarized in Tables \ref{tabmult} and \ref{tabdiv}. They show that the main effect of increasing (decreasing) the global intensity of the tidal dissipation is to increase (decrease) the migration timescale of the planet. The overall conclusions of the paper,  however, are not impacted by this tidal-dissipation efficiency effect. More specifically, the intensity of the effect of a given planet on the surface rotation rate of its central star is not impacted by a change in tidal-dissipation efficiency.}

{The observed properties of the WASP43 system are retrieved (in the case of a 0.7 $\rm M_{\odot}$ and a 2 $\rm M_{jup}$) at 3.0 Gyr when using the normal tidal dissipation evolution. When using the reduced efficiency, the age estimation increases  to 5.3 Gyr, which is expected given  that the tidal timescale is increased when the overall dissipation is  reduced. Interestingly, with the increased tidal dissipation, the age estimation is not shorter than that estimated using the normal dissipation, and is evaluated at 4.0 Gyr. With this increased dissipation, most of the close-in planets fall earlier into the central star. Since the main property of WASP-43 is that an orbiting planet is observed, these systems (i.e., the ones without a remaining planet) are discarded, and are no longer used for age estimation. As a consequence, the observed properties of the WASP-43 are retrieved at greater age compared to the normal tidal dissipation age.}



\section{Conclusion}

We showed that gyrochronology can be applied, with an error on the age estimation below 20\%, to massive close-in planetary systems where the orbital period of the massive planet is greater than $\rm P_{orb} \approx 4$ days (for a 1 $\rm M_{\odot}$ star and a planet less massive than 5 $\rm M_{jup}$). {In the range of initial planetary systems' conditions that we explored in this article, we highlighted that this limit increase with the stellar and planetary mass. }

{In this work, we determined that in the case of planetary engulfment, the effect of the transfer of angular momentum from the planetary orbit to the stellar rotation} could persist on a characteristic timescale of a few hundred Myr to a few Gyr (see Table \ref{tab1} and Fig. \ref{Deltat}). {We also investigated more dramatic events where the surface rotation rate of the star departs by more than 90\% from the rotation of the same but isolated star. We pointed out that in these scenarios the estimation of the age with the gyrochronology technique can lead to an age that is, in some cases, younger than several Gyr compared to the true age. Such massive departures can survive for tens of Myr  to a few hundred Myr after the event, which  most of the time is linked to a planetary engulfment, and can cause a more moderate but lasting 10\% departure for up to a few Gyr. {Provided the star experienced a planetary engulfment in its recent past, this effect could explain the bimodal age estimation of planetary systems such as GJ 504 \citep{Fuhrmann15,D'Orazi17,Bonnefoy18} for which two incompatible ages are found, depending on the technique employed}.}

Gyrochronology, if combined with other techniques, can be a convenient and rapid tool  for age determination; however, it should be kept in mind that the evolution of the stellar rotation could have been strongly modified by external torques such as star-planet tidal interaction.
To overcome this issue, we proposed a proof of concept of a new age determination technique that can be applied to such close-in planetary systems:  tidal-chronology. {This technique is based on the uniqueness of the path followed by a planetary system on the $\rm P_{\rm rot,\star}$(t)-$\rm P_{orb}$(t) plane.}

However, though an improvement over gyrochronology, the numerical and physical description of the tidal dissipation in stellar and planetary interior is currently not good enough to  use this age estimation alone; it  should be considered with caution and in combination with other age determination techniques. In this work we do not include the dissipation in the planet (which is still hardly theoretically constrained) nor the magnetic star-planet interaction \citep[see][]{Strugarek17}. The dissipation inside of the radiative core is also currently not physically described, and hence not numerically included. While the magnetic star-planet interaction torque is only present during the early PMS phase, the extra dissipations (from the planet, from the radiative core, and from taking into account the frequency response of the tidal dissipation) could shorten the migration timescale during the MS phase, and consequently reduce the estimated age of the system. The  behavior of these additional mechanisms is unfortunately not yet known, which prevents us from predicting their possible effect and impact on the conclusion reached in  Sect. \ref{gyro_appli}. However, while the quantitative results from Sect. \ref{gyro_appli} might change, we expect its general conclusion to remain the same. 


Within these limitations, we developed a promising technique that will benefit the community when all aspects of tidal and magnetic star-planet interactions are included in angular momentum evolution models.

\begin{acknowledgements}
{We thank the anonymous referee for the constructive comments about our work that increased the quality of the paper.} We warmly thank our colleagues Dr. Emeline Bolmont and Dr. Cilia Damiani for the helpful discussion about the orbital evolution of eccentric planets and the tidal dissipation inside giant planets. We thank Dr. Jacques Kluska for his help on statistical analysis, and Dr. Herv\'e Beust for the  discussions about the dynamical evolution of multiplanetary systems. F.G. acknowledges financial support from the CNES fellowship. This project has received funding from the European Research Council (ERC) under the European Union's Horizon 2020 research and innovation program (grant agreement No 742095; {\it SPIDI}: Star-Planets-Inner Disk-Interactions).  This work is supported by the French National Research Agency in the framework of the Investissements d'Avenir program (ANR-15-IDEX-02), through the funding of the ``Origin of Life'' project of the {Univ. Grenoble Alpes}.
\end{acknowledgements}

\bibliographystyle{aa} 
\bibliography{references} 

\begin{appendix}

\section{Exploration of different $|\rm \delta P|$?}
\label{explodeltap}

{Tables \ref{tab10}, \ref{tab25}, \ref{tab50}, and \ref{tab75} show the same data as in Table \ref{tab1}, but for different values of $|\rm \delta P|$= 10, 25, 50, and 75 \%.} 

\begin{table*}
\caption{Same as Table \ref{tab1}, but for $|\rm \delta P| = 0.1$.}
\label{tab10}
\begin{tabular}{|c|c|c|c|c|c|c|c|c|c|}
\hline
Mass & $\rm P_{init}$ & $\rm SMA_{init}$ &  $\rm Age_{10\%}$ & $\Delta t_{10\%}$  & $\rm P_{rot,10\%}$ & $\rm P_{orb,10\%}$ & $\rm P_{rot,isol.,10\%}$ & $\rm Age_{gyro}$ & $\rm Age_{gyro, isol.}$ \\
($M_{\odot}$) & (days) & (au) & (Myr) & (Myr) & (days) & (days) & (days) & (Myr) & (Myr) \\ 
\hline
\hline
0.5      &      6.00     &      0.0231   &      1902/102         &      2950/3600        &       13.13/1.19       &      0.2641/0.1666    &      22.26/2.78       &       575/\textit{8}   &      1474/\small\textit{36} \\
0.5      &      8.60     &      0.0196   &      6203/10  &      6000/4693        &       14.64/2.80       &      0.2642/0.1650    &      41.32/6.31       &       697/\textit{36}  &      4450/\small\textit{155} \\
\hline
\hline
0.7      &      5.00     &      0.0229   &      163/29   &      180/1724         &       6.04/1.30        &      0.2642/0.1581    &      7.08/1.97        &       165/\textit{11}  &      219/\small\textit{22} \\
0.7      &      6.00     &      0.0258   &      524/38   &      260/2116         &       9.03/1.27        &      0.2643/0.1651    &      10.99/2.38       &       338/\textit{10}  &      480/\small\textit{31} \\
0.7      &      7.00     &      0.0286   &      4155/35  &      8050/2770        &       17.94/1.30       &      0.2635/0.1630    &      36.71/2.92       &       1151/\textit{11}         &      4133/\small\textit{45} \\
0.7      &      8.00     &      0.0209   &      1405/14  &      2300/2991        &       15.32/3.36       &      0.2643/0.1621    &      19.36/4.74       &       869/\textit{58}  &      1318/107 \\
0.7      &      9.00     &      0.0226   &      3856/966         &      8050/3340        &       18.46/11.53      &      0.2639/0.1623    &      34.86/16.63      &       1211/523         &      3770/1006 \\
0.7      &      10.00    &      0.0242   &      7156/2656        &      5050/7650        &       18.99/13.37      &      0.2643/0.1607    &      47.04/27.76      &       1274/681         &      6437/2511 \\
\hline
\hline
1.0      &      4.00     &      0.0222   &      425/19   &      100/44   &       5.77/1.84        &      --/--    &      6.49/2.20        &      341/\textit{44}  &       421/\small\textit{61} \\
1.0      &      5.00     &      0.0258   &      1406/22  &      4250/904         &       12.22/1.01       &      --/--    &      13.57/2.98       &      1301/\textit{15}         &       1570/105 \\
1.0      &      6.00     &      0.0291   &      3957/1757        &      8250/6850        &       16.87/11.95      &      0.2623/0.2104    &      23.00/15.13      &       2316/1250        &      4027/1906 \\
1.0      &      7.00     &      0.0322   &      6058/3708        &      6150/8500        &       18.85/13.64      &      0.2606/0.2857    &      28.28/22.34      &       2822/1583        &      5825/3822 \\
1.0      &      8.00     &      0.0352   &      7459/5459        &      4750/6750        &       21.56/14.92      &      0.2530/0.3451    &      32.24/26.66      &       3588/1860        &      7360/5241 \\
1.0      &      9.00     &      0.0339   &      6710/4560        &      5500/7650        &       20.61/14.57      &      0.2480/0.3128    &      30.03/24.23      &       3311/1781        &      6484/4418 \\
1.0      &      10.00    &      0.0363   &      7711/5861        &      4500/6350        &       22.78/15.75      &      0.2562/0.3668    &      32.66/27.50      &       3958/2047        &      7532/5542 \\
1.0      &      11.00    &      0.0290   &      3312/1162        &      8900/7450        &       17.13/11.11      &      0.2599/--        &      20.60/12.26      &       2380/1097        &      3306/1310 \\
\hline 
\end{tabular}
\end{table*}

\begin{table*}
\caption{Same as Table \ref{tab1}, but for $|\rm \delta P| = 0.25$.}
\label{tab25}
\begin{tabular}{|c|c|c|c|c|c|c|c|c|c|}
\hline
Mass & $\rm P_{init}$ & $\rm SMA_{init}$ &  $\rm Age_{25\%}$ & $\Delta t_{25\%}$  & $\rm P_{rot,25\%}$ & $\rm P_{orb,25\%}$ & $\rm P_{rot,isol.,25\%}$ & $\rm Age_{gyro}$ & $\rm Age_{gyro, isol.}$ \\
($M_{\odot}$) & (days) & (au) & (Myr) & (Myr) & (days) & (days) & (days) & (Myr) & (Myr) \\ 
\hline
\hline
0.5      &      6.00     &      0.0231   &      1902/102         &      800/1150         &       13.13/1.19       &      0.2641/0.1666    &      22.26/2.78       &       575/\textit{8}   &      1474/\small\textit{36} \\
0.5      &      8.60     &      0.0196   &      6203/10  &      5200/1393        &       14.64/2.80       &      0.2642/0.1650    &      41.32/6.31       &       697/\textit{36}  &      4450/\small\textit{155} \\
\hline
\hline 
0.7      &      5.00     &      0.0229   &      223/29   &      80/514   &       5.77/1.30        &      --/0.1581        &      7.56/1.97        &       152/\textit{11}  &      246/\small\textit{22} \\
0.7      &      6.00     &      0.0258   &      564/38   &      160/626  &       8.87/1.27        &      --/0.1651        &      11.33/2.38       &       327/\textit{10}  &      506/\small\textit{31} \\
0.7      &      7.00     &      0.0286   &      4155/35  &      3550/1020        &       17.94/1.30       &      0.2635/0.1630    &      36.71/2.92       &       1151/\textit{11}         &      4133/\small\textit{45} \\
0.7      &      8.00     &      0.0209   &      1405/14  &      600/1091         &       15.32/3.36       &      0.2643/0.1621    &      19.36/4.74       &       869/\textit{58}  &      1318/107 \\
0.7      &      9.00     &      0.0226   &      3856/966         &      3300/1190        &       18.46/11.53      &      0.2639/0.1623    &      34.86/16.63      &       1211/523         &      3770/1006 \\
0.7      &      10.00    &      0.0242   &      7156/2656        &      5050/3450        &       18.99/13.37      &      0.2643/0.1607    &      47.04/27.76      &       1274/681         &      6437/2511 \\
\hline  
\hline 
1.0      &      4.00     &      0.0222   &      465/21   &      40/34    &       5.33/1.62        &      --/--    &      6.80/2.15        &      296/\textit{35}  &       457/\textit{59} \\
1.0      &      5.00     &      0.0258   &      1806/22  &      1650/76  &       12.39/1.01       &      --/--    &      15.59/2.98       &      1335/\textit{15}         &       2011/105 \\
1.0      &      6.00     &      0.0291   &      3957/1757        &      4550/3600        &       16.87/11.95      &      0.2623/0.2104    &      23.00/15.13      &       2316/1250        &      4027/1906 \\
1.0      &      7.00     &      0.0322   &      6058/3708        &      5450/5050        &       18.85/13.64      &      0.2606/0.2857    &      28.28/22.34      &       2822/1583        &      5825/3822 \\
1.0      &      8.00     &      0.0352   &      7459/5459        &      4750/5850        &       21.56/14.92      &      0.2530/0.3451    &      32.24/26.66      &       3588/1860        &      7360/5241 \\
1.0      &      9.00     &      0.0339   &      6710/4560        &      5500/5750        &       20.61/14.57      &      0.2480/0.3128    &      30.03/24.23      &       3311/1781        &      6484/4418 \\
1.0      &      10.00    &      0.0363   &      7711/5861        &      4500/6350        &       22.78/15.75      &      0.2562/0.3668    &      32.66/27.50      &       3958/2047        &      7532/5542 \\
1.0      &      11.00    &      0.0290   &      3612/1712        &      4700/3550        &       17.04/11.77      &      --/--    &      21.46/14.72      &      2356/1218        &       3557/1815 \\
\hline 
\end{tabular}
\end{table*}

\begin{table*}
\caption{Same as Table \ref{tab1}, but for $|\rm \delta P| = 0.50$.}
\label{tab50}
\begin{tabular}{|c|c|c|c|c|c|c|c|c|c|}
\hline
Mass & $\rm P_{init}$ & $\rm SMA_{init}$ &  $\rm Age_{50\%}$ & $\Delta t_{50\%}$  & $\rm P_{rot,50\%}$ & $\rm P_{orb,50\%}$ & $\rm P_{rot,isol.,50\%}$ & $\rm Age_{gyro}$ & $\rm Age_{gyro, isol.}$ \\
($M_{\odot}$) & (days) & (au) & (Myr) & (Myr) & (days) & (days) & (days) & (Myr) & (Myr) \\ 
\hline
\hline
0.5      &      5.00     &      0.0250   &      682/66   &      200/516  &       8.04/0.86        &      0.2642/0.1641    &      13.02/2.03       &       239/\textit{4}   &      566/\small\textit{21} \\
0.5      &      6.00     &      0.0231   &      1902/102         &      500/420  &       13.13/1.19       &      0.2641/0.1666    &      22.26/2.78       &       575/\textit{8}   &      1474/\small\textit{36} \\
0.5      &      8.60     &      0.0196   &      6203/10  &      1950/293         &       14.64/2.80       &      0.2642/0.1650    &      41.32/6.31       &       697/\textit{36}  &      4450/\small\textit{155} \\
\hline 
\hline 
0.7      &      5.00     &      0.0229   &      263/29   &      40/94    &       2.55/1.30        &      --/0.1581        &      7.88/1.97        &       \textit{35}/\textit{11}  &      265/\small\textit{22} \\
0.7      &      6.00     &      0.0258   &      624/38   &      80/86    &       7.72/1.27        &      --/0.1651        &      11.86/2.38       &       256/\textit{10}  &      549/\small\textit{31} \\
0.7      &      7.00     &      0.0286   &      4155/35  &      1550/110         &       17.94/1.30       &      0.2635/0.1630    &      36.71/2.92       &       1151/\textit{11}         &      4133/\small\textit{45} \\
0.7      &      8.00     &      0.0209   &      1655/17  &      250/80   &       13.11/2.82       &      --/--    &      21.19/4.38       &      657/42   &       1549/\textit{93} \\
0.7      &      10.00    &      0.0242   &      7156/2656        &      2800/1650        &       18.99/13.37      &      0.2643/0.1607    &      47.04/27.76      &       1274/681         &      6437/2511 \\
\hline 
\hline 
1.0      &      3.20     &      0.0234   &      40/12    &      4/18     &       1.54/1.17        &      --/--    &      2.78/1.89        &      \textit{32}/\textit{20}  &       92/\small\textit{46} \\
1.0      &      4.00     &      0.0222   &      485/29   &      20/22    &       4.04/1.87        &      --/--    &      6.95/2.91        &      181/\textit{46}  &       476/100 \\
1.0      &      5.00     &      0.0258   &      2106/22  &      400/12   &       10.82/1.01       &      --/--    &      17.01/2.98       &      1048/\textit{15}         &       2348/105 \\
1.0      &      6.00     &      0.0291   &      4307/2207        &      2250/1750        &       16.04/11.41      &      --/--    &      24.18/17.21      &      2115/1151        &       4404/2399 \\
1.0      &      7.00     &      0.0322   &      6058/3708        &      3150/3050        &       18.85/13.64      &      0.2606/0.2857    &      28.28/22.34      &       2822/1583        &      5825/3822 \\
1.0      &      8.00     &      0.0352   &      7509/5459        &      3400/3650        &       21.42/14.92      &      --/0.3451        &      32.44/26.66      &       3545/1860        &      7439/5241 \\
1.0      &      9.00     &      0.0339   &      6910/4560        &      3300/3550        &       19.96/14.57      &      --/0.3128        &      30.08/24.23      &       3127/1781        &      6501/4418 \\
1.0      &      10.00    &      0.0363   &      7961/5861        &      3550/3900        &       22.14/15.75      &      --/0.3668        &      33.39/27.50      &       3762/2047        &      7836/5542 \\
1.0      &      11.00    &      0.0290   &      4362/2262        &      1900/1450        &       15.53/11.08      &      --/--    &      23.59/17.13      &      1996/1093        &       4213/2379 \\
\hline 
\end{tabular}
\end{table*}

\begin{table*}
\caption{Same as Table \ref{tab1}, but for $|\rm \delta P| = 0.75$.}
\label{tab75}
\begin{tabular}{|c|c|c|c|c|c|c|c|c|c|}
\hline
Mass & $\rm P_{init}$ & $\rm SMA_{init}$ &  $\rm Age_{75\%}$ & $\Delta t_{75\%}$  & $\rm P_{rot,75\%}$ & $\rm P_{orb,75\%}$ & $\rm P_{rot,isol.,75\%}$ & $\rm Age_{gyro}$ & $\rm Age_{gyro, isol.}$ \\
($M_{\odot}$) & (days) & (au) & (Myr) & (Myr) & (days) & (days) & (days) & (Myr) & (Myr) \\ 
\hline
\hline
0.5      &      4.40     &      0.0229   &      722/46   &      120/136  &       6.85/1.03        &      --/--    &      12.49/1.81       &      180/\textit{6}   &       526/\small\textit{17} \\
0.5      &      5.00     &      0.0250   &      722/66   &      120/336  &       6.79/0.86        &      --/0.1641        &      13.31/2.03       &       177/\textit{4}   &      588/\small\textit{21} \\
0.5      &      6.00     &      0.0231   &      1952/102         &      350/340  &       12.32/1.19       &      --/0.1666        &      22.55/2.78       &       513/\textit{8}   &      1509/\small\textit{36} \\
0.5      &      8.60     &      0.0196   &      6203/10  &      950/213  &       14.64/2.80       &      0.2642/0.1650    &      41.32/6.31       &       697/\textit{36}  &      4450/\small\textit{155} \\
\hline
\hline
0.7      &      5.00     &      0.0229   &      263/39   &      20/46    &       2.55/0.97        &      --/--    &      7.88/1.91        &      \textit{35}/\textit{6}   &       265/\small\textit{21} \\
0.7      &      6.00     &      0.0258   &      644/38   &      60/60    &       6.50/1.27        &      --/0.1651        &      12.04/2.38       &       188/\textit{10}  &      564/\small\textit{31} \\
0.7      &      7.00     &      0.0286   &      4155/35  &      800/68   &       17.94/1.30       &      0.2635/0.1630    &      36.71/2.92       &       1151/\textit{11}         &      4133/\small\textit{45} \\
0.7      &      8.00     &      0.0209   &      1705/21  &      150/54   &       11.84/1.48       &      --/--    &      21.55/4.03       &      548/\textit{13}  &       1597/\small\textit{80} \\
0.7      &      9.00     &      0.0226   &      3856/1156        &      650/250  &       18.46/9.75       &      0.2639/--        &      34.86/18.00      &       1211/387         &      3770/1158 \\
0.7      &      10.00    &      0.0242   &      7156/2656        &      1550/800         &       18.99/13.37      &      0.2643/0.1607    &      47.04/27.76      &       1274/681         &      6437/2511 \\
\hline
\hline 
1.0      &      4.00     &      0.0222   &      491/31   &      14/20    &       1.70/1.99        &      --/--    &      7.00/3.50        &      \textit{38}/\textit{51}  &       482/139 \\
1.0      &      5.00     &      0.0258   &      2206/22  &      150/12   &       9.52/1.01        &      --/--    &      17.37/2.98       &      833/\textit{15}  &       2438/105 \\
1.0      &      6.00     &      0.0291   &      4757/2507        &      1050/950         &       14.07/10.34      &      --/--    &      25.13/18.49      &      1675/966         &       4718/2726 \\
1.0      &      7.00     &      0.0322   &      6608/4008        &      1700/1950        &       17.00/12.96      &      --/--    &      30.07/23.05      &      2346/1446        &       6498/4044 \\
1.0      &      8.00     &      0.0352   &      8109/5459        &      1850/2700        &       19.18/14.92      &      --/0.3451        &      34.00/26.66      &       2912/1860        &      8091/5241 \\
1.0      &      9.00     &      0.0339   &      7460/4810        &      1800/2400        &       18.32/14.07      &      --/--    &      32.12/24.84      &      2683/1674        &       7309/4618 \\
1.0      &      10.00    &      0.0363   &      8561/5911        &      1850/2900        &       20.26/15.64      &      --/--    &      35.68/27.52      &      3210/2023        &       8818/5547 \\
1.0      &      11.00    &      0.0290   &      4762/2512        &      650/550  &       13.72/10.12      &      --/--    &      24.46/17.91      &      1600/929         &       4493/2576 \\
\hline 
\end{tabular}
\end{table*}

\section{{Precision on $\rm P_{\rm rot,\star}$ and tidal-chronology degeneracy}}
\label{precdeg}

\begin{figure}[!ht]
\begin{center}
   \includegraphics[width=0.9\linewidth]{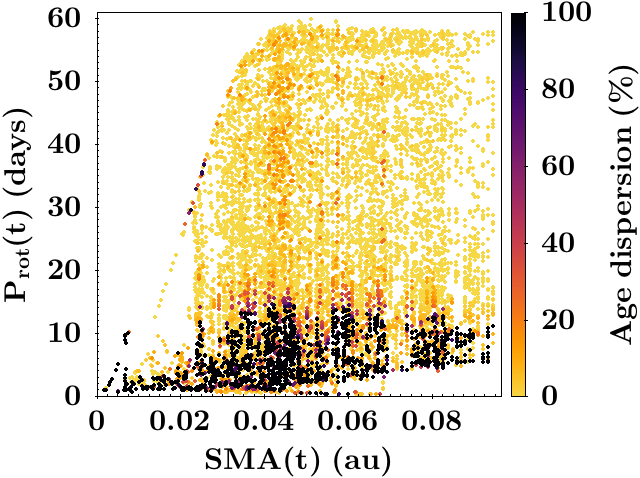} 
   \caption{{Same as Fig. \ref{Count}, but for  $\Delta \rm P_{\rm rot,\star} =0.1$ days.}}
\label{Count2}%
\end{center}
\end{figure}

\begin{figure}[!ht]
\begin{center}
   \includegraphics[width=0.9\linewidth]{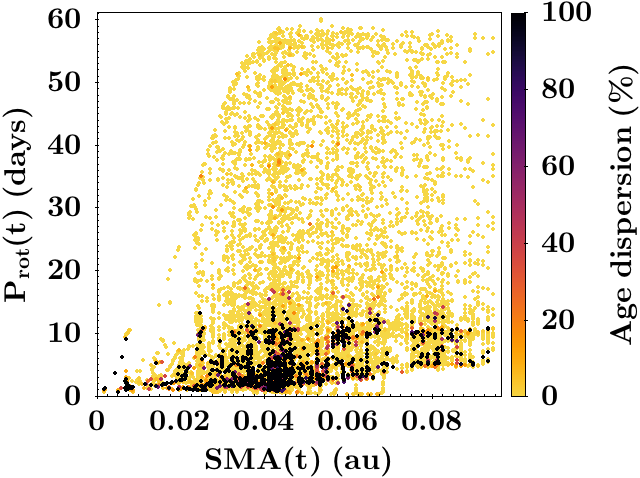} 
   \caption{{Same as Fig. \ref{Count}, but for  $\Delta \rm P_{\rm rot,\star} =0.01$ days.}}
\label{Count3}%
\end{center}
\end{figure}

{In this section we investigate the sensitivity of the degeneracies on the value of the error used for the $\rm P_{\rm rot,\star}$ measurements. In Figs. \ref{Count2} and \ref{Count3}, we used $\Delta \rm P_{\rm rot,\star} =0.1$ and $\Delta \rm P_{\rm rot,\star} =0.01$ , respectively. Naturally, we can see that the degeneracy increases when the precision on $\rm P_{\rm rot,\star}$ decreases. However, the degeneracy is still quite good even for the poorly constrained $\Delta \rm P_{\rm rot,\star} =0.1$ case.} 

\section{Tidal dissipation intensity and age estimation?}
\label{tidal_intensity}
\begin{table*}
 \begin{threeparttable}
\caption{{Error on the estimation of the age provided by the gyrochronology analysis in the case of $|\rm \delta P| = 0.5$. Here we consider a 0.7 $\rm M_{\odot}$ star around which orbits a 2 $\rm M_{jup}$ planet and for <$\mathcal{D}$>= <$\mathcal{D}$>$_{*100}$.  }}
\label{tabmult}
\begin{tabular}{|c|c|c|c|c|c|c|c|c|c|}
\hline
Mass & $\rm P_{init}$ & $\rm SMA_{init}$ &  $\rm Age_{50\%}$ & $\Delta t_{50\%}$  & $\rm P_{rot,50\%}$ & $\rm P_{orb,50\%}$ & $\rm P_{rot,isol.,50\%}$ & $\rm Age_{gyro}$ & $\rm Age_{gyro, isol.}$ \\
($M_{\odot}$) & (days) & (au) & (Myr) & (Myr) & (days) & (days) & (days) & (Myr) & (Myr) \\ 
\hline
\hline
0.7      &      6.00     &      0.0258   &      164/624  &      40/80    &       4.16/7.72        &      --/--    &      8.25/11.86       &      \textit{84}/256  &       288/549 \\
0.7      &      7.00     &      0.0286   &      965/4155         &      140/1550         &       10.05/17.94      &      --/0.2630        &      15.59/36.71      &       409/1151         &      895/4133 \\
0.7      &      8.00     &      0.0209   &      1655/1655        &      250/250  &       13.11/13.11      &      --/--    &      21.19/21.19      &      657/657  &       1549/1549 \\
0.7      &      9.00     &      0.0226   &      3856/3856        &      1350/1350        &       18.46/18.46      &      0.2639/0.2633    &      34.86/34.86      &       1211/1211        &      3770/3770 \\
0.7      &      10.40    &      0.0248   &      8707/8707        &      3400/3400        &       19.48/19.48      &      0.2642/0.2637    &      50.55/50.55      &       1333/1333        &      7319/7319 \\
\hline 
\end{tabular}
    \begin{tablenotes}
      \small
      \item  We adopted a dissipation that is two orders of magnitude higher <$\mathcal{D}$>$_{*100}$ than the frequency-averaged tidal dissipation used in this article. The listed data are for <$\mathcal{D}$>$_{*100}$ {(left)} and <$\mathcal{D}$> {(right)} and in the case of $|\rm \delta P| = 0.50$. We only show data that are in common between the two datasets.
    \end{tablenotes}
 \end{threeparttable}
\end{table*}

\begin{table*}
 \begin{threeparttable}
\caption{Same as Table \ref{tabmult}, but for a dissipation{ two orders of magnitude lower <$\mathcal{D}$>$_{/100}$}. }
\label{tabdiv}
\begin{tabular}{|c|c|c|c|c|c|c|c|c|c|}
\hline
Mass & $\rm P_{init}$ & $\rm SMA_{init}$ &  $\rm Age_{50\%}$ & $\Delta t_{50\%}$  & $\rm P_{rot,50\%}$ & $\rm P_{orb,50\%}$ & $\rm P_{rot,isol.,50\%}$ & $\rm Age_{gyro}$ & $\rm Age_{gyro, isol.}$ \\
($M_{\odot}$) & (days) & (au) & (Myr) & (Myr) & (days) & (days) & (days) & (Myr) & (Myr) \\ 
\hline
\hline
0.7      &      5.00     &      0.0229   &      3253/263         &      400/20   &       16.81/2.55       &      0.2644/--        &      32.98/7.88       &       1025/\textit{35}         &      3414/265 \\
0.7      &      6.00     &      0.0230   &      3854/884         &      550/40   &       17.57/4.97       &      0.2641/--        &      35.69/14.30      &       1109/116         &      3931/767 \\
0.7      &      7.20     &      0.0194   &      225/225  &      20/20    &       2.87/2.87        &      --/--    &      9.87/9.87        &      44/44    &       396/396 \\
0.7      &      8.00     &      0.0209   &      1755/1755        &      100/100  &       9.51/9.51        &      --/--    &      21.91/21.91      &      370/370  &       1645/1645 \\
0.7      &      9.00     &      0.0226   &      3906/3906        &      550/550  &       17.95/17.95      &      --/--    &      35.08/35.08      &      1152/1152        &       3812/3812 \\
0.7      &      10.40    &      0.0248   &      8707/8707        &      1500/1500        &       19.48/19.48      &      0.2642/0.2637    &      50.55/50.55      &       1333/1333        &      7319/7319 \\
\hline 
\end{tabular}
    \begin{tablenotes}
      \small
      \item The listed data are for <$\mathcal{D}$>$_{/100}$ {(left)} and <$\mathcal{D}$> {(right)}.
    \end{tablenotes}
 \end{threeparttable}

\end{table*}
{Tables \ref{tabmult} and \ref{tabdiv} show that when the efficiency of the tidal dissipation is globally reduced (increased) by a factor of 100, the star-planet tidal interaction timescale globally decreases (increases) compared to the normal tidal dissipation efficiency.}

\end{appendix}

\end{document}